\documentclass[aps,prb,twocolumn,showpacs,amssymb]{revtex4}
\usepackage{graphicx}
\usepackage{bm}% bold math

\begin{document}
%==============================================================================

\title{The origin of the stabilized simple-cubic structure in polonium}

\author {B. I. Min$^1$, J. H. Shim$^1$, Min Sik Park$^1$, Kyoo Kim$^1$, 
S. K. Kwon$^1$, and S. J. Youn$^2$} 

\affiliation{$^1$Department of Physics, Pohang University of Science 
        and Technology, Pohang 790-784, Korea \\
             $^2$Department of Physics Education, 
        Gyeongsang National University, Jinju 660-701, Korea}

\date{\today}

\begin{abstract}

The origin of the stabilized simple-cubic (SC) structure in Po 
is explored by using the first principle band calculations.
We have found that the prime origin is the inherent strong spin-orbit (SO) 
interaction in Po, which suppresses the Peierls-like structural instability
as usually occurs in $p$-bonded systems. 
Based on the systematic analysis of electronic structures, 
charge densities, Fermi surfaces, and susceptibilities of 
Se, Te, and Po, we have proved that the stable crystal structure 
in VIA elements is determined by the competition between the SO splitting 
and the crystal field splitting induced by the low-symmetry 
structural transition. The trigonal structure is stabilized in Se and Te by
the larger crystal field splitting than the SO splitting, 
whereas in Po the SC structure is stabilized by the large SO splitting.

\end{abstract} 

\maketitle

%==============================================================================

%\section{Introduction}
%\label{sec:intro}

Among the elements in the periodic table,
only Po (polonium) is known to crystallize in the simple-cubic (SC) 
structure in nature \cite{beamer}. 
Radioactive Po, which was discovered by Pierre and Marie Curie in 1898,
requires difficult sample preparation, and
so not many experimental and theoretical studies have been carried out.
Po, which is located in the VIA column of the periodic table,
has a valence electron configuration of $6s^26p^4$ ($Z$=84),
and exhibits two metallic phases of $\alpha$ and $\beta$
which are formed in the SC and the 
rhombohedral (trigonal) structure below and above $\sim$ 348$K$,
respectively \cite{max,des}.
Note that the isoelectronic elements located in the same VIA column,
Se (selenium) and Te (tellurium), crystallize in the trigonal 
structure (space group $P3_{1}21$) with the helical chain 
arrangements of atoms, which run parallel to 
the crystallographic $c$ axis of the hexagonal setting.
Hence the coordination number is two in trigonal Se and Te 
in contrast to six in simple-cubic Po (SC-Po). The ratio of the
intrachain ($d_1$) bond length to the interchain ($d_2$) bond length 
is $\frac{d_2}{d_1} = 1.50, 1.23$, and $1.0$ for Se, Te, and Po, 
respectively. The coordination number of two in trigonal Se and Te can be
understood based on the simple octet rule \cite{Gaspard}.

The reason why metallic Po has a stable SC structure
has not been fully addressed yet.
By using the relativistic parameterized extended H\"{u}ckel method,
Lohr \cite{lohr} has found a hint that the helical structure 
as a distortion of a SC structure might be 
quenched in the case of Po due to its very large spin-orbit (SO) coupling.
However, their calculation was not self-consistent and
adopted the atomic value of the SO coupling parameter. 
They did not study the structural energetics.
Recently, using the {\it ab-initio} pseudopotential band method,
Kraig {\it et al.} \cite{kraig} showed that 
the SC structure is preferred by Po to face-centered or
body-centered cubic structure.
They argued that the large $s$-$p$ splitting in Po would produce
the stable SC structure. However, they could not
explain why this happens only in Po but
not in other elements with similarly large $s$-$p$ splittings.
Moreover, the SO interaction was not taken into account in their
band calculations. Lach-hab {\it et al.} \cite{Lach} studied
the structural energetics for Po using the tight-binding (TB) band method.
After having determined the TB parameters by fitting the TB bands
to the semirelativistic full-potential linearized augmented plane-wave 
(FLAPW) \cite{wim} bands, they took account of the 
SO effect {\it a posteriori} by employing the atomic value of the
SO coupling parameter.
They found that the SC structure is the most stable among
the close-packed structures for both cases
with and without the SO effect included. They did not 
consider the lower symmetry structures with the internal atomic relaxation,
as observed in Se and Te. 

%-----------------------------------------
\begin{figure}[b]
\includegraphics[scale = 0.33,angle=270]{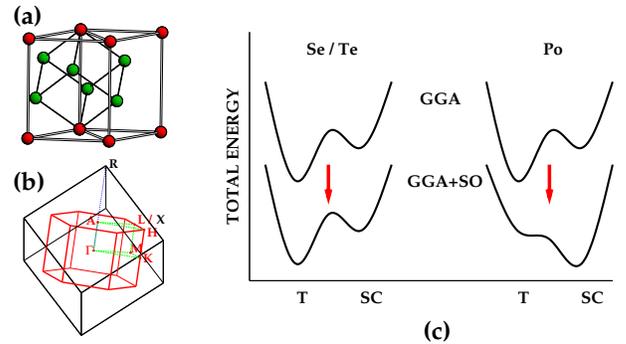}
\caption{(a) A hexagonal unit cell with tripled volume of 
a SC unit cell. 
(b) The hexagonal Brillouin zone (BZ) compared to the SC BZ.
Note that $\Gamma$-A and $\Gamma$-L in the hexagonal BZ correspond to 
$\frac{1}{3}\Gamma$-R and $\Gamma$-X in the SC BZ.
(c) Schematic total energy curves for Se, Te, and Po in the hypothetical
structural space.
T and SC represent trigonal and simple-cubic structures,
respectively.
Both in the GGA and GGA+SO, trigonal Se and Te are more stable.
For Po, trigonal Po is more stable in the GGA, while
SC-Po becomes more stable in the GGA+SO.
}
\label{fig1}
\end{figure}
%-----------------------------------------

%-----------------------------------------
\begin{figure}[t]
\includegraphics[scale = 0.55]{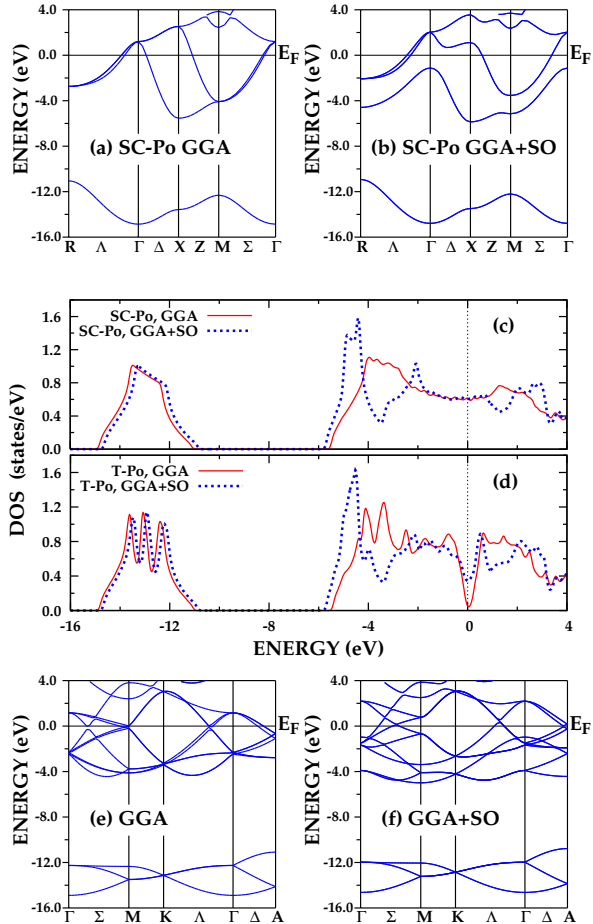}
\caption{(a)-(b) The GGA and GGA+SO band structures of SC-Po.
(c)-(d) DOSs of SC-Po and trigonal Po. 
The latter corresponds to the stable phase
obtained after the geometry optimization in the GGA scheme.
In each frame, DOSs obtained by the GGA and GGA+SO schemes 
are compared.
(e)-(f) The GGA and GGA+SO band structures of SC-Po drawn in the
frame of the hexagonal BZ. There are several band crossings 
at $E_F$ in the GGA, but not in the GGA+SO.
}
\label{fig2}
\end{figure}
%-----------------------------------------

A systematic crystallographic transformation can occur from the 
SC lattice to the trigonal lattice.
As shown in Fig.~\ref{fig1}(a),
the SC lattice can be described by a hexagonal lattice
in which the hexagonal $a$ axis corresponds to the face diagonal
and the hexagonal $c$ axis to the body diagonal of cube.
Then the size of the unit cell becomes tripled 
with $c/a=\sqrt\frac{3}{2}$. The trigonal structure of $\beta$-Po
corresponds to the elongated SC structure simply along the body diagonal
direction. In the trigonal structures of Se and Te, the internal atomic 
position parameters are also changed in addition to elongation \cite{Kresse}. 

To explore the origin of stabilized SC structure in Po,
we have investigated the electronic and structural properties of Se, Te, and
Po systematically by using the first-principles FLAPW band method
incorporating the SO interaction. 
For the calculations, the QMD-FLAPW \cite{wim} and 
the WIEN2K \cite{Blaha} codes are utilized. 
The band structure results obtained from 
both codes are qualitatively coincident with each other.
Below we will use mainly WIEN2K results, 
since the local orbital basis is implemented in WIEN2K in order
to describe the relativistic $6p_{1/2}$ wave function 
correctly \cite{Singh}.
The exchange-correlation interaction is treated by
the generalized gradient approximation (GGA)\cite{Perdew,LDA}.

We have first studied the structural energetics by optimizing
the atomic positions for both trigonal
and SC structures of Se, Te, and Po 
with (GGA+SO) and without (GGA) the SO interaction included.
In the GGA scheme, the trigonal structures are
more stable than SC structures by 0.092, 0.031, and 0.004 eV/atom
for Se, Te, and Po, respectively.
The GGA+SO scheme, however, yields different results.
The incorporation of the SO coupling reduces the stability of the 
trigonal structure.
For Se and Te, the trigonal structure is still more stable
than the SC structure by 0.090 and 0.021 eV/atom, respectively.
In contrast, for Po, the SC structure 
becomes more stable than the trigonal structure by 0.016 eV/atom. 
Thus the total energy behaves as in Fig.~\ref{fig1}(c).
There are two local minima in the hypothetical structural space,
the trigonal (T) and the SC structure.
The SO coupling effect does not change the global minima 
in Se and Te, but changes that in Po from trigonal Po to SC-Po.
This finding indicates that the SO interaction plays an essential
role in stabilizing the SC structure of Po.

Indeed band structures in Fig.~\ref{fig2}(a)-(b) reveal that the effect of
the SO interaction is substantial for SC-Po, as compared to that for Se and Te.
The lowest bands in Fig.~\ref{fig2}(a)-(b)
correspond to Po $6s$ states, while the bands near the Fermi level ($E_F$) 
to Po $6p$ states.
Due to the large energy gap between $6s$ and $6p$ states,
there will not be a considerable mixing between them,
and thus Po can be classified as a $p$-bonded system.
When the SO interaction is taken into account, 
the degenerate $6p$ bands along $\Gamma$-X, $\Gamma$-M, and $\Gamma$-R are 
split so that the doubly degenerate 
$6p_{1/2}$ band is separated from the upper half-filled
$6p_{3/2}$ band.
Due to the SO splitting, the small hole Fermi surface at $\Gamma$ disappears.
The SO splitting of $p$ states at $\Gamma$ 
for SC-Po is as much as 3.16 eV, which is much larger
than those for SC structures of Se and Te, 0.53 and 1.05 eV.

In Fig.~\ref{fig2}(c)-(d), we have compared the density of states (DOS) 
of SC-Po and that of trigonal Po, which are obtained 
by the GGA and GGA+SO schemes.
Trigonal Po here corresponds to the stable phase determined by the 
geometry optimization in the GGA scheme.
Notable in the GGA-DOS is the pseudogap feature at $E_F$ 
for trigonal Po, which would give rise to the semi-metallic behavior. 
The pseudogap appears between the nonbonding lone-pair
and the antibonding Po $6p$ state,
because of the splitting between the bonding
and antibonding Po $6p$ state in the chain arrangement of trigonal 
Po \cite{joan}. 
Hence the DOS at $E_F$, $N(E_F)$, becomes much reduced in trigonal Po 
from that in SC-Po.  In fact, the stable trigonal phases of Se and Te
result from this gap formation near $E_F$.
In the case of Se, the energy splitting between the bonding
and the antibonding state is so large that
the system becomes insulating, while Te is semimetallic.

In the GGA+SO scheme, the $6p_{1/2}$ state is separated out, 
and so the band shape and N($E_F$) are changed a lot. 
Occupied band widths are broadened with more
weight at low energy side, which produces energy gains 
for both SC-Po and trigonal Po.
The stable SC-Po in the GGA+SO scheme reflects that 
the SO induced energy gain is larger for SC-Po than for trigonal Po.
It is because $N(E_F)$ for trigonal Po is rather enhanced 
producing some energy loss.
Note that, for Se and Te,  both the GGA and GGA+SO schemes
yield similar DOS, implying no noticeable energy gain from the SO interaction.
These results suggest that the stable crystal structure
in VIA elements is mainly determined by the
competition between the SO splitting and the crystal field splitting 
due to the low symmetry structure.
That is, for Se and Te, the larger crystal field splitting 
between the intra-chain bonding and antibonding states 
stabilizes the trigonal structure,
whereas, for Po, the larger SO splitting stabilizes the SC structure.

The stable trigonal structure of Se and Te can be understood in terms of
the three dimensional (3D) Peierls distortion of the SC structure
\cite{Gaspard,Kresse,Burdett}.
In general, $p$-bonded systems would favor the SC structure to maximize the 
directional $p_{\sigma}$ bonding. 
However, three mutually orthogonal linear chains in the SC structure
will be easily distorted by the Peierls mechanism 
when the linear chain is partially filled. 
In the case of Se and Te, each chain is 2/3-filled, 
and so the energy gain can be achieved by the trimerization with 
the short-long-long 
bond alternation. Decker {\it et al.} \cite{Decker} recently
demonstrated this mechanism for Te by comparing the band structures of
the undistorted SC and the distorted trigonal structure.
The structural distortion from the SC to the trigonal structure
induces a splitting of degenerate bands near $E_F$, which yields
the energy gain for trigonal Te.

%-----------------------------------------
\begin{figure}[t]
\includegraphics[scale = 0.4, angle=270]{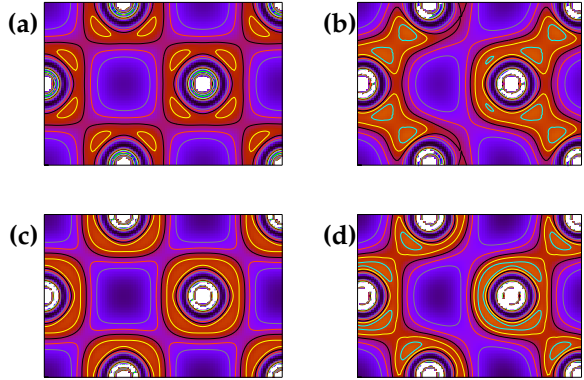}
\caption{The occupied charge density distributions 
of Po $6p$ bonding states in the energy range of $-6.0 \sim -3.0$ eV.
(a) SC-Po (GGA), (b) trigonal Po (GGA), (c) SC-Po (GGA+SO), 
and (d) trigonal Po (GGA+SO).
}
\label{fig3}
\end{figure}
%-----------------------------------------

%-----------------------------------------
\begin{figure}[t]
\includegraphics[scale = 0.37, angle=270]{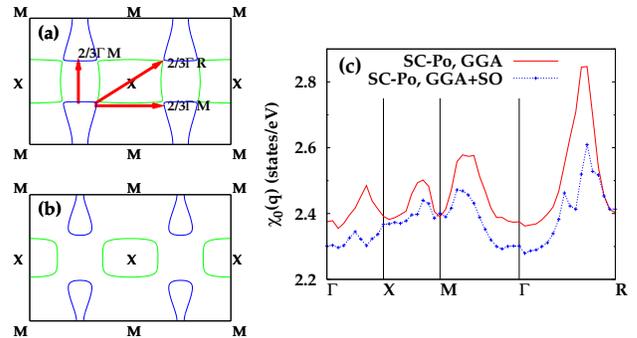}
\caption{Fermi surfaces of SC-Po derived from the third (green) 
and fourth (blue) bands in the GGA (a) and the GGA+SO (b).
(c) Susceptibilities of SC-Po in the GGA and the GGA+SO.
}
\label{fig4}
\end{figure}
%-----------------------------------------

We have examined the above scenario for SC-Po.
Figure~\ref{fig2}(e)-(f) display the GGA and GGA+SO band structures 
of undistorted SC-Po which are projected in the hexagonal BZ.
The GGA band of Fig.~\ref{fig2}(e) shows several band crossings at $E_F$ 
which are created by the back-folding of bands of
the larger SC BZ into the smaller hexagonal BZ (see Fig.~\ref{fig1}(b)).
Then, as in Te \cite{Decker}, a Peierls-like structural distortion would
split the band crossings at $E_F$ to yield the energy gain.
In contrast, the GGA+SO band of Fig.~\ref{fig2}(f) shows the band crossings 
not at $E_F$ but at higher or lower energy side far apart from $E_F$.
This is because of the substantial band shift by the SO splitting.
Accordingly, no energy gain would result from the Peierls-like structural 
distortion.
This proves how the SO interaction stabilizes the SC structure of Po.
When Decker {\it et al.} \cite{Decker} demonstrated
this mechanism for Te, they did not take into account the
SO interaction. The SO induced band shift in Te, however, is not 
large enough to suppress the Peierls-like instability.

The bonding characters can be investigated by the charge density plot.
Figure~\ref{fig3} shows the occupied charge density distributions 
of Po $6p$ bonding states in the energy range of $-6.0 \sim -3.0$ eV
for both SC-Po and trigonal Po.
They are plotted on the special plane of hexagonal unit cell
to see the bonding character along the helical zig-zag chains 
in the $c$-direction.  The GGA charge density for SC-Po
(Fig.~\ref{fig3}(a)) shows larger bonding character between Po atoms 
than the GGA+SO charge density (Fig.~\ref{fig3}(c)).
This implies that the directional bonding between atoms becomes weakened
when incorporating the SO interaction, and so the charge density becomes 
more isotropic.
This feature is more clearly seen in trigonal Po which manifests
the prominent chain nature.
The GGA charge density for trigonal Po (Fig.~\ref{fig3}(b))
shows a pronounced intrachain bonding character along the zig-zag chains.
In the GGA+SO charge density (Fig.~\ref{fig3}(d)),
the SO interaction weakens the intrachain directional bondings 
so that the anisotropic chain nature is suppressed.
This is consistent with the enhanced $N(E_F)$ in the
GGA+SO DOS in Fig.~\ref{fig2}(d).
The charge density plots reveal that the SO interaction weakens 
the directional bondings between Po atoms and so suppresses 
the 1D chain nature and the corresponding Peierls-like structural instability.

The structural transition can also be studied by analyzing the 
behavior of the charge susceptibility $\chi_0(q)$.
$\chi_0(q)$'s in Fig.~\ref{fig4}(c) are obtained from the band structures
of SC-Po.
$\chi_0(q)$ in the GGA scheme shows the highest peak near
$\vec{Q} =\frac{2}{3}$$\Gamma$-R, indicating a possible structural instability 
at this $\vec{Q}$. 
This $\vec{Q}$ vector is coincident with the reciprocal lattice vector
connecting the hexagonal BZ boundaries including A (see 
Fig.~\ref{fig1}(b)), at which the bands of SC-Po
along $\Gamma$-R are folded back.
Therefore, this peak is closely related to the trigonal 
distortion of the SC structure along the (111) direction.
Noteworthy in Fig.~\ref{fig4}(c) is the substantial 
reduction of this peak of $\chi_0(q)$ when incorporating the SO interaction.
The absence of the structural instability in Po
would be closely correlated to this behavior of $\chi_0(q)$.
One can also note the peaks in $\chi_0(q)$ 
near $\vec{q} =\frac{2}{3}$$\Gamma$-X, $\frac{2}{3}$$\Gamma$-M,
$\frac{2}{3}$X-M. These peaks and the peak near $\vec{Q}$
are expected to arise from the quasi-1D Fermi surfaces of SC-Po,
which are formed by the third and fourth bands in Fig.~\ref{fig2}(a).
Namely, each $\vec{q}$ producing the peak in $\chi_0(q)$ corresponds to the 
nesting vector of the quasi-1D Fermi surfaces projected in each symmetry plane.
The GGA Fermi surfaces in Fig.~\ref{fig4}(a), which are drawn in the (220)
plane of the SC BZ, exhibit the clear quasi-1D nature
with the corresponding nesting vectors.
As shown in Fig.~\ref{fig4}(b), the SO interaction breaks 
these quasi-1D Fermi surfaces into pieces, and so reduces the nesting effect.
This situation in Po is different from that in Se and Te, 
in which the 1D nature is preserved even with the SO interaction, so that 
the nesting effect is still active.
 
Finally, it is worthwhile to discuss the stable structures of 
neighboring elements of Po, Bi (Z=83) and 
At (Z=85), both of which are also $p$-bonded metallic systems 
with the large SO interaction.
In nature, Bi crystallizes in the trigonal ($\alpha$-arsenic) 
structure \cite{Shick}. As for radioactive At, no structural information is 
available because of its too short half-life of only 8.3 hours.
$\alpha$-arsenic structure of Bi can be understood as arising from
the Peierls-like distortion, as discussed above \cite{Burdett,Gaspard}. 
Since the $6p$-bands in Bi are half-filled, the tendency of the
Peierls instability is predominating so as to drive the structural distortion
despite the large SO interaction.
In the case of At, the $6p$-bands are 5/6-filled so that
the SO interaction would dominate over the Peierls instability
as in Po.
Hence the SC structure is expected to be stabilized for At.

In conclusion, we have clarified that the origin of the stabilized 
SC structure in Po is its inherent strong SO interaction.
The large SO interaction in Po weakens the directional bondings between atoms 
so as to suppress the Peierls-like distortion.
Our study also reveals that the stable crystal structure in VIA elements is 
determined by the competition between the SO splitting and the crystal 
field splitting induced by the structural transition to a lower symmetry.
For Se and Te, the larger crystal field splitting between the intra-chain
bonding and the antibonding state stabilizes the trigonal structure,
whereas, for Po, the larger SO splitting stabilizes 
the SC structure.

%Acknowledgments $-$
This work was supported by the SRC program of MOST/KOSEF
and in part by the KRF.


\begin{thebibliography}{99}

   \bibitem{beamer} W. H. Beamer and C. R. Maxwell, 
           J. Chem. Phys. {\bf 14}, 569 (1946).
   \bibitem{max} C. R. Maxwell, J. Chem. Phys. {\bf 17}, 1288 (1949).
   \bibitem{des} R. J. Desando and R. C. Lange, 
            J. Inorg. Nucl. Chem. {\bf 28}, 1837 (1966).
 \bibitem{Gaspard} J.-P. Gaspard, A. Pellegatti, F. Marinelli, C. Bichara C,
	 Phil. Mag. B {\bf 77}, 727 (1998).
   \bibitem{lohr} L. L. Lohr, Inorg. Chem. {\bf 26}, 2005 (1987).
   \bibitem{kraig} R. E. Kraig, D. Roundy, and M. L. Cohen,
            Solid State Comm. {\bf 129}, 411 (2004).
   \bibitem{Lach} M. Lach-hab, B. Akdim, D.A. Papaconstantopoulos, M.J. Mehl, 
	N. Bernstein, J. Phys. Chem. Solid {\bf 65}, 1837 (2004).
   \bibitem{Kresse} G. Kresse, J. Furthmuller, and J. Hafner,
	Phys. Rev. B {\bf 50}, 13181 (1994).
   \bibitem{wim} E. Wimmer, H. Krakauer, M. Weinert, and A. J. Freeman,
            Phys. Rev. B {\bf 24}, 864 (1981);
	M. Weinert, E. Wimmer, and A. J. Freeman, {\it ibid.} {\bf 26}, 
	4571(1982); H. J. F. Jansen and A. J. Freeman {\it ibid.} {\bf 30},
	561 (1984).
 \bibitem{Blaha} P. Blaha, K. Schwarz, G. K. H. Madsen, K. Kvasnicka, 
	and J. Luitz, WIEN2K (Karlheinz Schwarz, Technische Universitat Wien,
	 Austria, 2001).
 \bibitem{Singh} D. Singh, Plane waves, Pseudopotentials,
	 and the LAPW Method (Kluwer Academic, Boston, 1994).
 \bibitem{Perdew} J.P. Perdew and Y. Wang, 
            Phys. Rev. B {\bf 45}, 13244 (1992).
 \bibitem{LDA} We have found that the GGA describes the ground states of
	Se, Te, and Po better than the LDA.
 \bibitem{joan} J. D. Joannopoulos, M. Schl\"{u}ter, and M. L. Cohen,
            Phys. Rev. B {\bf 11}, 2186 (1975).
 \bibitem{Burdett} J. K. Burdett and S. Lee, J. Amer. Chem. Soc.,
	  {\bf 105}, 1079 (1983).
 \bibitem{Decker} A. Decker, G. A. Landrum, and R. Dronskowski,
	Z. Anorg. Allg. Chem., {\bf 628}, 295 (2002).
 \bibitem{Shick} A. B. Shick, J. B. Ketterson, D. L. Novikov, 
	and A. J. Freeman, Phys. Rev. B {\bf 60}, 15484 (1999).

%            Nature {\bf 395}, 677 (1998).
%            Appl. Phys. Lett. {\bf 74}, 1737 (1999).
%            Phys. Rev. Lett. {\bf 74}, 1171 (1995).
%            Solid State Comm. {\bf 124}, 77 (2002).
%            Rev. Mod. Phys. {\bf 40} 790 (1968).
%            Physica C {\bf 341-348}, 785 (2000).

\end{thebibliography}
\end{document}